# Reviewing energy system modelling of decentralized energy autonomy[1]

Jann Michael Weinand[1], Fabian Scheller[2], Russell McKenna[2]

[1] Chair of Energy Economics, Karlsruhe Institute of Technology, Hertzstraße 16, 76187 Karlsruhe, Germany

[2] DTU Management, Technical University of Denmark, 2800 Kgs. Lyngby, Denmark

Corresponding author: Jann Michael Weinand, jann.weinand@kit.edu, +49 721 608 44444

**Abstract**

Research attention on decentralized autonomous energy systems has increased exponentially in the past three decades, as demonstrated by the absolute number of publications and the share of these studies in the corpus of energy system modelling literature. This paper shows the status quo and future modelling needs for research on local autonomous energy systems. A total of 359 studies are roughly investigated, of which a subset of 123 in detail. The studies are assessed with respect to the characteristics of their methodology and applications, in order to derive common trends and insights. Most case studies apply to middle-income countries and only focus on the supply of electricity in the residential sector. Furthermore, many of the studies are comparable regarding objectives and applied methods. Local energy autonomy is associated with high costs, leading to levelized costs of electricity of 0.41 $/kWh on average. By analysing the studies, many improvements for future studies could be identified: the studies lack an analysis of the impact of autonomous energy systems on surrounding energy systems. In addition, the robust design of autonomous energy systems requires higher time resolutions and extreme conditions. Future research should also develop methodologies to consider local stakeholders and their preferences for energy systems.

**Highlights**

- Literature review includes 123 studies about local autonomous energy systems.
- Mainly simulation or optimization with a central planner perspective
- The most common models HOMER and EnergyPlan should be extended in future studies
- Levelized costs of electricity for local energy autonomy are 0.41 $/kWh on average.
- Future work should focus on non-technical dimensions, open models and data

---

**Abbreviations**

| Abbreviations | Meaning |
| --- | --- |
| ABS | Abstract |
| ar | Article |
| BBO | Biogeography Based Optimisation |
| CHP | Combined heat and power |
| Doctype | Type of document, e.g. article or review |
| EA | Energy autonomy |
| ES | Energy system |
| EV | Electric vehicle |
| HIC | High income county |
| KEY | Keyword |
| LCOE | Levelized cost of electricity |
| LEA | Local energy autonomy |
| LIC | Low income country |
| LMIC | Lower middle income country |
| LPSP | Loss of load probability |
| MCDA | Multi-criteria-decision analysis |
| PV | Photovoltaics |
| RE | Renewable energy |
| UMIC | Upper middle income country |

## 1. Introduction

Between 1993 and 2017, the percentage of the worldwide population with access to electricity increased from 77% to around 89% [1]. Since 2012, more than 100 million people per year have gained access to electricity. However, it is estimated that even in 2030 about 670 million people will still have no access to electricity [2]. Most people without access live in rural areas (84%) and in sub-Saharan Africa or developing Asia (95%) [3]. Negative examples include the developing countries of Burundi, Chad and Malawi, where less than 15% of the population have access to electricity [1]. Many of the newly electrified regions in developing countries apply off-grid solutions with diesel engines due to long distances to the national grid. Such completely energy autonomous systems are able to meet the energy demands of an entire community without energy imports [4].

Whereas these completely autonomous (i. e. off-grid) energy systems (ESs) exist in developing countries mainly due to cost considerations, there are also efforts by municipalities and regions to become energy autonomous in industrialized countries with complete electrification (i.e. grid-connected). This is due to the energy transition and the related environmental awareness [5] as well as the desire of citizens to play an active role in energy supply and to be less dependent on central markets and structures (e. g. [6,7]). The majority of municipalities with energy autonomy (EA) aspirations strive for balanced EA and the focus is usually on electrical energy [5]. In this context, ESs are balanced autonomous if they are energy neutral, i.e. the annual locally provided energy exceeds the annual demand [4]. In contrast to a completely energy autonomous solution, imports and exports are possible.

Local energy autonomy (LEA) can take different forms and degrees. In particular, autonomous ESs need to exhibit the basic criteria as defined by Rae and Braedly [4]. First, the local system is able to generate at least as much energy to meet the demands. Second, the local system allows energy shifting possibilities for times in which there is a temporal mismatch between demand and supply (i.e. through storage or in the case of balanced autonomy through energy infrastructure). Third, the system is capable of operating independently on a *stand-alone* or *off-grid* basis. Thereby, local autonomy efforts are related to active participation of the community or rather the system components are owned by the community members. In general, autonomy efforts are directly related to the notion of *self-governance* and *community ownership* [4]. In this study, EA is defined the same way: it focusses on plants inside the municipality which tend to be operated by the local community, and may also include conventional technologies. However, due to the dependency on fuel transport and the high costs of diesel-based ESs, a supply consisting at least in part of renewable energies (REs) could be worthwhile in these cases. Figure 1 shows a schematic representation of decentralized (autonomous) energy systems.

As the above examples demonstrate, balanced or completely autonomous ESs are related to different objectives and have different effects on the local setting but possibly also on the overarching system. Due to the increasing relevance of LEA (c.f. section 3), there is a need to elaborate and define transition process aspects and successful transition pathways. Therefore, the aim of this study is to review the literature on LEA, i.e. in villages, districts, municipalities and regions, in order to identify the current state of the art and gaps or starting points for future studies. This spatial resolution is chosen since similar conditions apply, for example from a technical point of view (decentralised energy technologies) but also from a social point of view, such as the number and type of stakeholders. Studies on individual buildings and larger regions such as entire nations are therefore excluded. In contrast to the multitude of existing reviews, this paper for the first time shows a comprehensive overview and quantitative analysis of applied methods in EA case studies at the local level (cf. section 2).

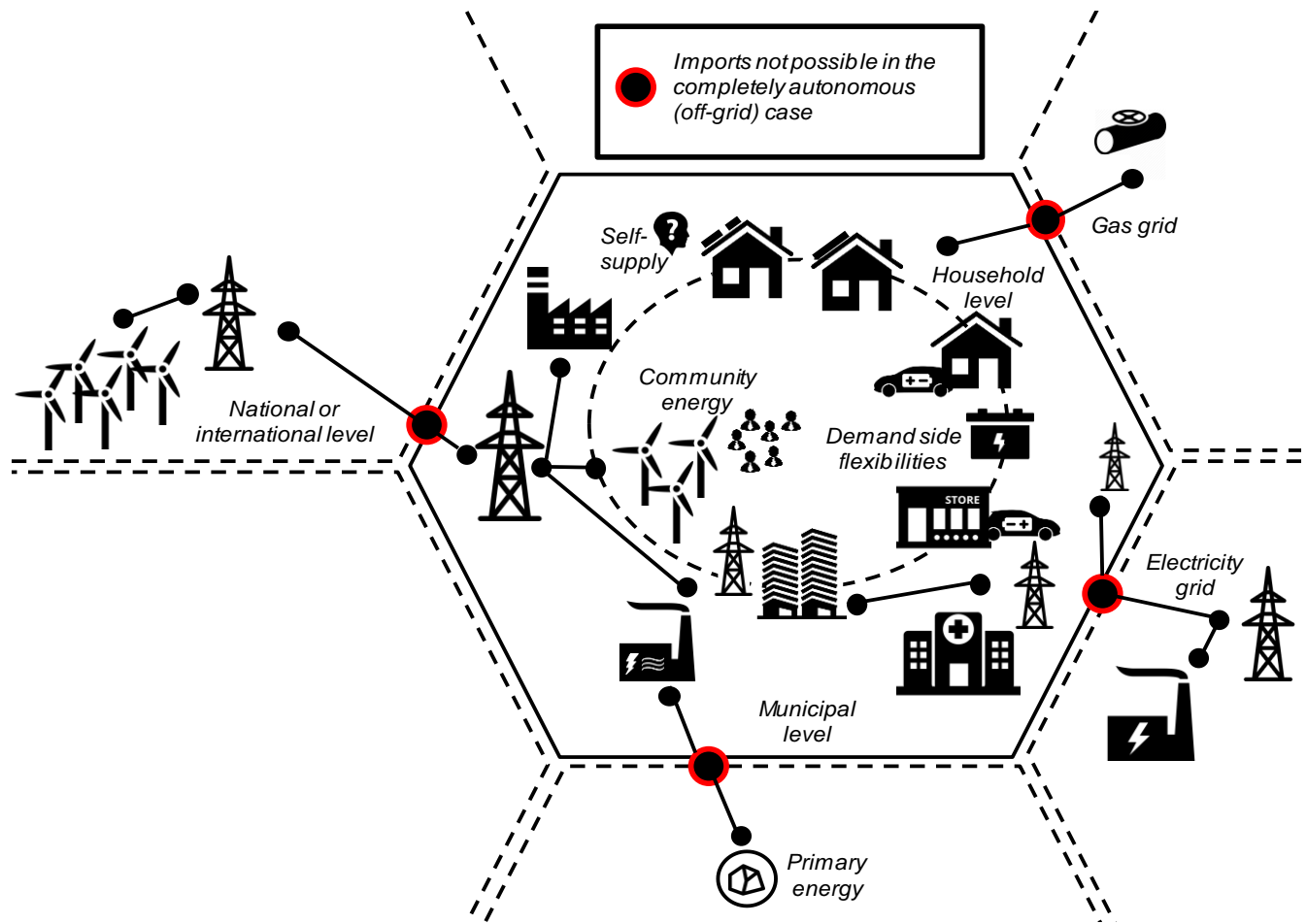

*Figure 1: Schematic representation of a decentralized energy system. Technologies, which are occuring in the studies, can not be comprehensively demonstrated. In this review only studies on the municipal level and regional level are included, i.e. no studies on the household or national level. In contrast to the completely energy autonomous case (off-grid), imports from the national level are permitted in the case of balanced energy autonomy.*

This paper is structured as follows: section 2 gives an overview of already published literature reviews about ES analyses with a focus on decentralized EA. Subsequently, section 3 presents the methodology for the literature search. Section 4 then presents and discusses the most important findings from the analysis of the literature, before the studies are critically assessed in section 5. A summary and conclusions are given in section 6.

**2. Existing literature reviews**

Some review papers have already examined studies on local ES planning for larger (e.g. urban areas) [8,9] and smaller regions (e.g. municipalities / communities) [4,10–16]. In some cases, the focus is on some form of EA [4,10,15,16], or this topic is at least briefly discussed [11,14]. A few other review papers [17–20] discuss EA, but do not focus on the local level as defined in section 1.

Even though there are some undeniable differences regarding the motives of LEA efforts in developed and developing countries, due to the growing utilization of decentralized RE generators LEA projects also represent a business opportunity in industrialized countries according to the review of Engelken et al. [21]. Despite all of the criticism voiced by Heard et al. [19], the feasibility and viability of such local RE systems have been demonstrated in various studies as shown in the reviews of Brown et al. [17] and Hansen et al. [18].

In this context, a strategy to achieve this feasibility requires the discussion of crucial factors regarding EA. Kaundinya and Ravindranath [16] present both general success and failure stories of corresponding autonomy projects. According to the authors, a generalized approach to assess the suitability of off-grid and grid-connected systems, based on techno-economic-financial-environmental feasibility does not find adequate coverage in the literature yet. Issues and trends shaping local ESs are conceptually summarized by Keirstead et al. [8] and Koirala et al. [11]. Concerning this matter, local ESs fit very well into the neo-liberal ideas of self-reliance and independence [11]. A shift towards local autonomous systems, however, is not only associated with a host of social, financial and environmental benefits. Key challenges are the degree and scale of EA, the matching of demand with supply, the importance of socio-economic and political factors and the structural requirements in remote communities [11].

These results go in line with the findings of the review of Rae and Bradley [4]. While taking into account the different drivers of balanced or completely autonomous energy projects, the review investigates environmental, political, economic, technical and social concerns. Even though generation and utilization of RE is primarily a technical challenge, social and political aspects are the most important factors in its implementation in the community. The conceptual framework of EA by Müller et al. [15] shares a similar focus: the involvement and motivation of administrations and community stakeholders are decisive for the successful transition towards LEA.

Additionally, ensuring stable supply and reliability against all plausible outcomes in RE availability might raise cost and complexity of the systems due to the impacts of worst-case conditions [19]. Distributed energy generation as with local autonomous projects encompasses a wide variety of technologies which tend to be highly sensitive to the deployment context [14]. Thus, it is important to consider the necessary spatial and temporal boundaries of the region or community.

Some of the mentioned aspects have also been posed in terms of future modelling needs. Trans-disciplinary and multi-dimensional features of low-carbon community model approaches are outlined by Nakata et al. [20]: models should consider e.g. the utilization of waste for energy, the inclusion of various sectors, and approaches related to energy-for-development issues in rural areas of developing countries. Dependent on the objectives and constraints, Gamarra and Guerrero [10] point out innovative planning guidelines by reviewing optimization techniques applied to microgrid planning. While the microgrid siting problem of autonomous systems requires robust methodologies, the operation of the autonomous system is only possible with reliable energy management systems. Thereby, stochastic optimization could be one solution for a more realistic estimate [16]. Besides, Scheller and Bruckner [12] present requirements for ES modelling at the municipal level and discuss existing optimization models concerning their fundamental approaches. They provide future modelling needs for successful ES analyses which are also linked with the mentioned challenges of autonomy projects.

In contrast to many other literature reviews, the present paper presents a comprehensive overview and quantitative analysis of applied methods in case studies on energy autonomous systems at the local level. Of the reviews discussed in this section, only [4,10,11,14–16] focus on EA at the local level. Of these six papers, only Kaudinya et al. [16] and Gamarra et al. [10] concentrate on methodological aspects. However, the studies differ from the present review study by concentrating on the comparison of grid-connected and off-grid ESs [16] or only on microgrid studies [10]. Furthermore, the studies are from 2009 [16] and 2015 [10], respectively, and since then, the published articles on LEA have more than doubled (cf. section 3). Apart from the general topic, this review also covers new aspects such as a compilation of costs for local energy autonomous systems (cf. section 4.7).

## 3. Review methodology

The literature on local ES analysis has increased exponentially from 1990 until 2019 (cf. Figure 2), along with the LEA efforts described in section 1. Scopus[2] was used for the primary literature search, since it covers a wider range of journals [22] as well as more recent sources [23] than other databases like *Web of Science*. The *Initial search* query in Table 1 results in a total of 2,453 studies (cf. Figure 2). The search query contains the methodology (e. g. ES analysis or simulation), the spatial resolution (e. g. municipality or region) and the restriction that it is a peer-reviewed article. 359 (15%) of the 2,453 studies on local ES analyses deal with autonomous ESs and have also exponentially increased in recent years (*Adjusted search* in Table 1 and Figure 2).

*Table 1: Different search queries for the literature search in Scopus. The abbreviation "TITLE-ABS-KEY" is used to search for the terms in the title, abstract and keywords of the article. Included are original research articles (cf. "(DOCTYPE, "ar")" in the search query), which were published between 1990 and 2019.*

| Search name | Search query | Number of studies |
|---|---|---|
| Initial search | TITLE-ABS-KEY ("energy system" AND ("simulation" OR "modelling" OR "optimisation" OR "analysis") AND ("region" OR "municipalities" OR "municipality" OR "communities" OR "community" OR ("district" AND NOT "district heating") OR "city" OR "cities" OR "town" OR "remote")) AND (LIMIT-TO (DOCTYPE,"ar")) | 2,453 |
| Adjusted search | TITLE-ABS-KEY ("energy system" AND ("simulation" OR "modelling" OR "optimisation" OR "analysis") AND ("region" OR "municipalities" OR "municipality" OR "communities" OR "community" OR ("district" AND NOT "district heating") OR "city" OR "cities" OR "town" OR "remote") AND ("off-grid" OR "off grid" OR ("100%" AND "RE") OR ("100%" AND "renewable") OR "100%-renewable" OR ("energy" AND "autonomy") OR ("energy" AND "autarky") OR ("energy" AND "self-sufficiency") OR ("energy" AND "self-sufficient") OR "energy independent" OR "stand-alone" OR "energy autonomous" OR "island system")) AND (LIMIT-TO (DOCTYPE,"ar")) | 359 |
| "Energy system analysis" search | TITLE-ABS-KEY("energy system" AND ("simulation" OR "modelling" OR "optimisation" OR "analysis")) AND (LIMIT-TO(DOCTYPE,"ar")) | 12,368 |

The increasing importance of the topics could be only related to the generally exponentially increasing number of publications. However, the share of local ES analyses in the field of ES analysis (*Energy system analysis search* in Table 1) has increased from 8% (1990) to 20% (2019) and that of local energy autonomous systems from 0% to 3%.

The 359 studies about LEA were examined for suitability for this review. 236 studies were excluded for the reasons outlined in Table 2. For 122, most of them were excluded because of an unsuitable spatial resolution (e.g. ES analysis of a single building). A total of 123 studies remained ([24–105,105–132,132–140,140–146]), which were mainly published in the journals Energy and Renewable Energy (cf. Table 3). In addition, Table 4 shows the ten most globally cited articles on LEA.

---
[2] https://www.scopus.com/search/form.uri?display=basic

*Table 2: Studies that resulted from the adjusted search in Scopus and are not considered in this literature review for the reasons given in the table.*

| Exclusion criterium | Number of studies | References |
|---|---|---|
| The study does not consider energy autonomy as defined in section 1 (i. e. at least balanced autonomy has to be analysed) | 30 | [147–176] |
| Autonomy is considered in a different context than energy | 2 | [177,178] |
| Autonomy is only mentioned as a future target in the paper | 3 | [179–181] |
| The spatial resolution of the study does not match our definition of local energy systems (cf. section 1) | **122** | |
| • Single consumer / households / building | 41 | [182–222] |
| • Single commercial application | **57** | |
|     o Agricultural well | 2 | [223,224] |
|     o Desalination unit | 7 | [225–231] |
|     o Cellular base station / telecommunication unit | 11 | [232–242] |
|     o Hospital / healthcare facility | 5 | [243–247] |
|     o Hotel | 5 | [248–252] |
|     o Library | 1 | [253] |
|     o Wireless sensor nodes | 1 | [254] |
|     o Machinery laboratory | 1 | [255] |
|     o Agricultural application (farm or irrigation area) | 6 | [256–261] |
|     o Voter registration centre | 1 | [262] |
|     o Desert safari camp | 1 | [263] |
|     o Touristic facility | 1 | [264] |
|     o Charging station | 1 | [265] |
|     o Mining site | 3 | [266–268] |
|     o Factory / enterprise | 3 | [269–271] |
|     o Refinery | 1 | [272] |
|     o Road lighting system | 1 | [273] |
|     o University facility / school | 4 | [274–277] |
|     o Clean water and toilet system | 1 | [278] |
|     o Wastewater treatment plant | 1 | [279] |
| • Large regions | 3 | [280–282] |
| • One or several countries | 21 | [283–303] |
| Analysis of a single energy plant / technology | 35 | [304–338] |
| Aerospace applications | 2 | [339,340] |
| Climate analyses | 4 | [341–344] |
| Study focusses on control strategies of an energy system | 13 | [345–357] |
| Study introduces a new model without autonomy case study | 3 | [358–360] |
| Study develops load profiles for off-grid areas | 2 | [361,362] |
| Study focusses on qualitative analysis | 15 | [363–377] |
| Analysis of a given 100 % renewable system | 2 | [378,379] |
| Text language: Korean | 2 | [380,381] |
| Publication not found | 1 | [382] |

*Table 3: Distribution of the studies among the journals in which they were published. Only those journals are shown which have published five or more studies.*

| Journal | Number of studies | Share in 123 studies [%] |
|---|---|---|
| Energy | 16 | 13 |
| Renewable Energy | 14 | 12 |
| Energy Conversion and Management | 9 | 7 |
| International Journal of Renewable Energy Research | 9 | 7 |
| Applied Energy | 7 | 6 |
| Energies | 7 | 6 |
| Journal of Cleaner Production | 6 | 5 |
| Solar Energy | 5 | 4 |

*Table 4: Most relevant articles on local energy autonomy, based on global citations (15.01.2020).*

| Article | Title | Journal | Global citations |
|---|---|---|---|
| Ashok 2007 [33] | Optimised model for community-based hybrid energy system. | Renewable Energy | 295 |
| Ma et al. 2014 [134] | A feasibility study of a stand-alone hybrid solar-wind-battery system for a remote island. | Applied Energy | 257 |
| Kanase-Patil et al. 2010 [72] | Integrated renewable energy systems for off grid rural electrification of remote area. | Renewable Energy | 187 |
| Østergaard and Lund 2011 [100] | A renewable energy system in Frederikshavn using low-temperature geothermal energy for district heating. | Applied Energy | 180 |
| Ma et al. 2014 [85] | Technical feasibility study on a standalone hybrid solar-wind system with pumped hydro storage for a remote island in Hong Kong. | Renewable Energy | 158 |
| Maleki and Askarzadeh 2014 [88] | Optimal sizing of a PV/wind/diesel system with battery storage for electrification to an off-grid remote region: A case study of Rafsanjan, Iran. | Sustainable Energy Technologies and Assessments | 102 |
| Haghighat Mamaghani et al. 2016 [60] | Techno-economic feasibility of photovoltaic, wind, diesel and hybrid electrification systems for off-grid rural electrification in Colombia. | Renewable Energy | 95 |
| Gupta et al. 2010 [58] | Steady-state modelling of hybrid energy system for off grid electrification of cluster of villages. | Renewable Energy | 92 |
| Rohani and Nour 2014 [111] | Techno-economical analysis of stand-alone hybrid renewable power system for Ras Musherib in United Arab Emirates. | Energy | 89 |
| Askarzadeh and dos Santos Coelho 2015 [36] | A novel framework for optimization of a grid independent hybrid renewable energy system: A case study of Iran. | Solar Energy | 85 |

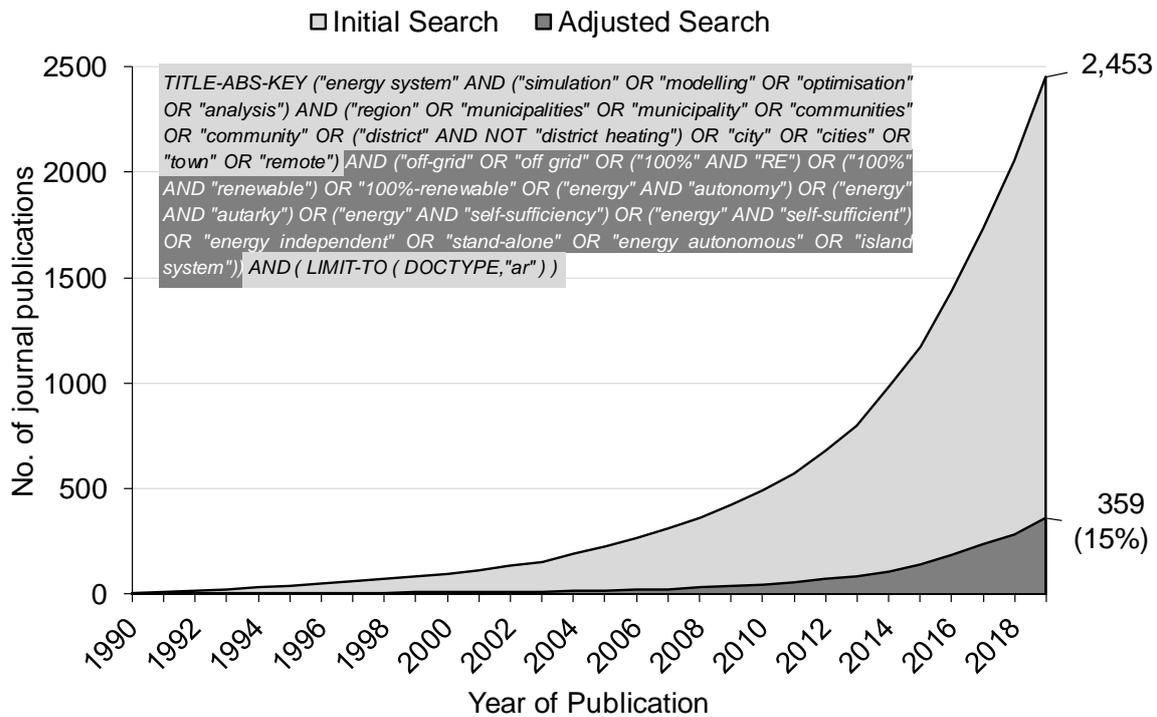

*Figure 2: Development of the number of journal publications for two search queries in Scopus. The bright curve contains the publications on energy system analyses at the local level. For the dark curve, the search query has been adjusted so that the studies also deal with autonomous energy systems. The last search in Scopus was performed on November 20, 2019.*

## 4. Results of the literature review

This section presents and discusses the main findings of the analysis of the 123 studies. First, section 4.1 explains the methods and models used in the studies. Subsequently, section 4.2 shows the system boundaries of the case studies. The type and feasibility of EA under consideration is presented in section 4.3. In the following section 4.4, the temporal resolution is discussed. Section 4.5 then indicates the included technologies before section 4.6 examines the consideration of grid infrastructures. Finally, section 4.7 compares and discusses the costs for LEA resulting in the studies.

The supplementary Microsoft Excel file in the online appendix contains all the information discussed in this article. Not all aspects could be identified in all studies. In some of these cases, the information in the Excel file are given in parentheses, which indicates assumptions based on detailed analysis of the study. If, in the following sections, the shares in the number of studies do not add up to 123 or the percentages do not add up to 100%, this is due to the fact that not all information could be retrieved from every study.

### 4.1. Methodologies and models

The methods employed in the reviewed literature (cf. Table 5) range from simple energy balancing calculations (e. g. [70,99,104]) to simulations (e. g. [79,94,102]), metaheuristics (e. g. genetic algorithm [69], discrete harmony search [88], artificial bee swarm optimization [89,90] or flower pollination algorithm [113]), mixed-integer linear optimizations (e. g. [30]) and multi-objective optimizations (e. g. [112]).

*Table 5: Overview of the general methodologies applied in the studies.*

| General methods | Study | Number of studies |
|---|---|---|
| Artificial bee swarm optimization | [89,90] | 2 |
| Biography based optimization algorithm | [133] | 1 |
| Chaotic search | [128,129] | 2 |
| Cukoo search | [115] | 1 |
| Discrete harmony search | [47,88,128] | 3 |
| Electricity System Cascading Analysis | [64] | 1 |
| Energy balancing calculation | [70,99,104,122] | 4 |
| Firefly algorithm | [114] | 1 |
| Flower Pollination algorithm | [113] | 1 |
| Fuzzy analytic hierarchy process | [98] | 1 |
| Genetic algorithm | [69,106] | 2 |
| Grey relation analysis | [98] | 1 |
| Life cycle cost analysis | [87] | 1 |
| Multi-objective particle swarm optimization | [37,125,126] | 3 |
| Multi-objective crow search algorithm | [68] | 1 |
| Multi-objective optimization | [112] | 1 |
| Non-dominated sorting genetic algorithm-II | [125,126] | 2 |
| Optimization | [30,34,49,54,56,58,59,63,72,74,78,92,98,109,117,124,146] | 16 |
| Multi-criteria-decision analysis (MCDA) | [80,92] | 2 |
| Particle swarm optimization | [36,61,101] | 3 |
| Simulated annealing | [128] | 1 |
| Simulation | [24–29,31–33,35,38–43,45,46,48,50–53,55,57,60,62,65–67,71,75–77,79–86,91,93–97,100,102,103,105,107,108,110,111,116,118–121,123,127,130–132,134–145] | 69 |

When investigating the methods, it is striking that in most cases only simplified calculations are carried out. For example, the number and size of plants is usually predefined and not optimized. Simulations are frequently used (56%, cf. Table 5) and often referred to as optimization. The best examples are the simulation softwares HOMER and EnergyPLAN, which are used in 61 (50%) and 7 (6%) studies respectively (cf. Table 6). Apart from HOMER and EnergyPLAN, other models appear only once or twice or are not specified in the studies (cf. Table 6).

Although most HOMER studies call it optimization, in its core HOMER is a simulation model [383]. The simulation is used to balance energy generation and consumption and calculate the costs. Thousands of scenarios with different parameters can be simulated in a sensitivity analysis. Subsequently, in the so-called optimization, the best solution is selected from among all scenarios depending on the selected criterion (e.g. minimization of costs or fuel usage) [383]. This explorative approach to identifying a pareto front does not necessarily yield the optimal solution. The vast majority of the studies reviewed here in which HOMER is used have a similar structure: First, the economic parameters, the load profile as well as the renewable potentials and the ES under consideration are described for a particular application. The best ES is then usually selected on the basis of costs (97% of cases). These studies therefore typically focus on case studies rather than methodological innovations. The aim of these studies is usually to reduce the diesel consumption of a remote off-grid area. In three studies, newly developed methods were compared with the HOMER model. The results showed that a Biogeography Based Optimization (BBO) algorithm [133], a Genetic Algorithm [69] or the so-called LINGO model [72] perform better than the HOMER model in terms of computing time and minimization of costs. The BBO algroithm, for example, found a better solution than HOMER and reduced the computing time from 15 h to 0.7 h [133].

*Table 6: Models used in the analysed literature.*

| Model | Study | Number of studies |
|---|---|---|
| HOMER or HOMER Pro | [24–29,32,35,38,39,41–43,45,46,48,50,53,55,57,60,62,65,66,71,75–77,80–83,86,91,93,95–97,105,107,108,110,111,118–121,127,131,132,134–142,144,145] | 61 |
| EnergyPLAN | [44,51,52,100,116,123,130] | 7 |
| RE³ASON | [92,124] | 2 |
| BeWhere / Phasma | [117] | 1 |
| FINE / TSAM | [146] | 1 |
| FlexiGIS | [30] | 1 |
| H$_2$RES | [79] | 1 |
| IREOM | [73] | 1 |
| ISLA | [98] | 1 |
| LINGO v.10 | [72] | 1 |
| OSeMOSYS | [56] | 1 |
| P²IONEER | [78] | 1 |
| RegFin | [104] | 1 |

Despite the weaknesses that HOMER shows in determining an optimal ES, the advantages of the tool should also be appreciated. It is an open access tool that can be used by everyone and does not require a lot of computing power. Therefore, the model is particularly useful for studies on remote areas in countries with low or middle income (75% of the HOMER studies reviewed here) according to the country classification of the World Bank [384] (cf. Table 8).

In contrast to the HOMER model, the EnergyPLAN model was only used for case studies in countries with middle (China [130]) or high (Italy [44], Croatia [51,116], Denmark [52,100], Ireland [123]) income. However, this model is also classified "as a simulation tool rather than an optimisation tool" [385] and only includes a dispatch optimisation. This means that the user has to specify the technologies and thus has to have a comprehensive understanding of ES analysis (as in HOMER). A step in the right direction was the introduction of Homer Pro, which is intended to simplify the use of the HOMER software for inexperienced users [386].

Beyond these two models, there are more and more model approaches that generate inputs for optimization models on the basis of publicly available data such as OpenStreetMap[3]. Examples in the reviewed literature are the FlexiGIS model [30] and the RE³ASON model [92,124]. At least for FlexiGIS an open-source publication in GitHub is planned, according to the authors [30]. These models would enable inexperienced users in the field of ES analysis to determine the energy potentials of a region and optimize the ES. Neither potentials nor technologies and plant sizes would have to be determined before application of the models.

### 4.1.1. Perspective

In most ES analyses, a *central planner* was used as the perspective. Only in Ramchandran et al. [107], the perspective of a Renewable Energy Service Company was taken instead. On the one hand, these central planner approaches show the macroeconomic optimum. On the other hand, however, these approaches often fall short of realisation: it could be difficult to convince individual homeowners to install the technologies in their homes that are optimal from a macroeconomic but not a business point of view. Therefore, studies of possible incentive systems that could encourage homeowners to implement the macroeconomic EA plan are required. First conclusions could be drawn by comparing the optimal ES at building level from the perspective of the building owner on the one hand and from the macroeconomic perspective on the other.

Promising approaches include for example the combination of optimization approaches with *multi-criteria-decision analysis* (MCDA), as in [80] and [92]. These types of analyses do not yet include the perspective of every inhabitant, but at least the perspective of their most important representatives. This could help to strengthen the feasibility of an EA plan.

### 4.1.2. Target criterion

The adopted perspective is closely linked to a further shortcoming of many of the studies examined: the choice of the target criterion. In 103 (84%) of the studies only costs were minimized. As already shown in section 1, however, costs are not the only reason for EA projects. In a few studies, a different target criterion than costs is applied, but these analyses do not represent an improvement since also only one criterion is considered: technical feasibility [85,116], annual efficiency [102,103] or coal consumption [130]. However, beside the above-mentioned MCDA approaches, other multi-objective approaches represent improvements. In addition to costs, the criteria in Table 7 are also taken into account in the reviewed studies. Possible weightings for the criteria could either be determined on site before the case studies, or taken from surveys such as in [5].

---

[3] https://www.openstreetmap.org/

*Table 7: Applied target criteria besides cost in the studies with MCDA or multi-objective optimizations.*

| Criterion | Study | Number of studies |
|---|---|---|
| $CO_2$ emissions | [70,80,92,98,112] | 5 |
| Land use | [54,70,80,83] | 4 |
| Security of supply | [37,68,83,98] | 4 |
| Forecast accuracy | [125,126] | 2 |
| Renewable share in energy supply | [37,83] | 2 |
| Social acceptance | [80,98] | 2 |
| Community net imports | [92] | 1 |
| Creation of jobs | [80] | 1 |
| Ease of installation and operation | [98] | 1 |
| Flexibility of the system for future expansion | [98] | 1 |
| Gross domestic production | [104] | 1 |
| Health issues | [80] | 1 |
| Human development index | [80] | 1 |
| Noise | [80] | 1 |
| Risk of flash floods | [80] | 1 |
| Technical efficiency | [80] | 1 |
| Technical reliability | [80] | 1 |
| Technical lifespan | [80] | 1 |
| Technical scalability | [80] | 1 |
| Technical maturity | [80] | 1 |
| Universal education and gender equality | [80] | 1 |
| Water consumption | [54] | 1 |
| Water quality | [80] | 1 |

## 4.2. System boundaries

This section first persents the spatial resolution and location of the case studies in the reviewed literature (cf. section 4.2.1). Section 4.2.2 then shows that the focus is primarily on the demand product electricity. Finally, the demand sectors considered in the studies are highlighted (cf. section 4.2.3).

### 4.2.1. Spatial resolution and location

Table 8 classifies the case studies of the reviewed papers according to the income classification of *The World Bank* [384]. Most of the studies were conducted in the *lower middle income country* (LMIC) India (21), *upper middle income countries* (UMIC) Iran (17) and China (6) as well as the *high income country* (HIC) Germany (7). Whilst the case studies in India and Iran focused mainly on the ESs of remote areas without grid connection, in Germany a complete electrification already exists. The case studies on complete autonomous ESs in HICs are therefore more about isolating communities from the transmission grid. These studies are linked to the question whether the ES transformation should be achieved through decentralised or centralised expansion of RE sources. In [387], for example, the decentralised expansion is evaluated as more cost-effective for the German case due to higher required transmission grid expansion costs in the centralised case. The possible impacts of autonomous communities on the surrounding ES (cf. section 4.3) therefore plays a very important role especially in the HIC studies.

Whilst the *low income countries* (LIC), the LMIC and the UMIC mainly consider remote rural areas as case studies with 100%, 92% and 74% of the studies, respectively, these remote areas account for only 23% of the studies on HIC. Instead, studies on HIC also often investigate EA for islands (23%), cities (20%), regions (17%) and municipalities (14%). This is also reflected by the number of residents examined in the case studies: in the studies on LIC, LMIC and UMIC case studies with a maximum of 4,750 inhabitants are analysed; in the HIC studies, case studies with up to 640,000 inhabitants are investigated. The exact area or city names of the case studies as well as the number of households and inhabitants examined can be found in the supplementary Excel file.

In this context, it is noticeable that the complexity of the applied methodology adapts to the size of the considered region. This means, the smaller a region is chosen, the more details can be included in the analysis. For example, in Waenn et al. [123] the operation of the ES with 640,000 inhabitants is determined with the help of a less complex simulation (EnergyPLAN). The largest case study in which an ES is designed with the help of an optimization is in Schmidt et al. [117]: a large region with 21,000 inhabitants. However, in Schmidt et al. only two time slices are considered during the optimization (cf. section 4.4.2) to reduce the model complexity.

In [60,76,104,106,111,124,144] different spatial scales were compared as case studies, i.e. the number of households or inhabitants was varied. However, these studies do not yet provide indications about the optimal size of energy autonomous ESs.

*Table 8: Classification of the countries according to The World Bank [384], in which the case studies are conducted in the reviewed literature as well as the share in the total of 123 publications. This country classification is based on the gross national income per capita [384].*

| **Income group [384]** | **Gross national income per capita [384] [$]** | **Countries and studies** | **Share [%]** |
|---|---|---|---|
| Low income | [0; 1,025) | Ethiopia [66,96], Rwanda [144], Tanzania [140], Yemen [31] | 4 |
| Lower middle income | (1,025; 3,995] | Bangladesh [50,91,136], Cameroon [95,127], Egypt [113], Ghana [26,28], India [29,33,45–47,58,59,72,73,77,81,87,93,101,106,107,109,114,115,131,133], Indonesia [32,120], Nigeria [27,137,139], Pakistan [71,75,110], Philippines [98], Timor-Leste [56] | 33 |
| Upper middle income | (3,995; 12,375] | Algeria [43,135], Brazil [118], China [69,80,82–84,130], Colombia [60], Cuba [132], Iran [25,34–36,41,61,63,65,68,88–90,125,126,128,129,141], Iraq [138], Malaysia [62,64,119,121], Maldives [55], Mexico [54], Turkey [53] | 30 |
| High income | (12,375; inf) | Australia [40,57], Austria [117], Canada [38,39,105,122,145], Croatia [51,79,116], Denmark [52,100], Finland [104], Germany [30,70,78,92,94,124,146], Greece [102,103], Hong Kong [85,86,134], Ireland [123], Italy [44], Japan [143], Korea [76], Oman [24], Saudi Arabia [42,108,142], Scotland [48], Sweden [37], Switzerland [112], United Arab Emirates [97,111], USA (Alaska) [49] | 33 |

### 4.2.2. Demand

In all the studies reviewed, the electricity demand of the ES is included. Heating or cooling demand, on the other hand, is only considered in 30 (24%) or 13 (11%) of the studies, respectively. As already indicated in section 1 for EA projects, this also demonstrates the focus on electricity in the literature. In most cases, the demand is based on time series that have been determined or collected beforehand. However, there are also examples of EA case studies such as [30,92,124], in which the demand and load profiles are determined automatically on the basis of publicly available data. The electricity and heat generation technologies used in the studies are presented in section 4.5.

In addition to electricity and heat, other demand products such as food are also indirectly taken into account, for example through land-use competition as in Schmidt et al. [117]. However, only direct demand products are discussed in the present literature review. This includes the demand for water considered in ten studies. In [28,46,47,63,93] this is considered by the electricity demand of a water pump, e. g. for an agricultural well. In [67,74,79,89] the ES contains a desalination unit for water distillation. In Fuentes-Cortés et al. [54] the *water-energy nexus* is considered in the analysis, which means that water demand in the energy supply is taken into account. In this case the water demand includes fresh water for households, water used for regulating the temperature of the thermal demand as well as water needed as by-product in the combined heat-and-power (CHP) units. Water consumption is included in [54] alongside costs and land use in the multi-objective function of the optimization model. Therefore, [54] in particular shows a suitable way to consider water demand in future studies about autonomy. The studies [47,54,74,79] are the only examples which consider all three types of demand (electricity, heat and water).

### 4.2.3. Consumption sectors

Among the consumption sectors, mainly the residential sector is considered (102 studies; 83%), followed by the commercial sector (55; 45%), industrial sector (23; 19%) and transportation sector (11; 9%). The ES is usually designed for all considered sectors. By contrast, in Bagheri et al. [38] the residential, commercial and industrial sectors are examined in separate analyses. Thereby, the ES for the industrial sector shows the lowest *levelized cost of electricity* (LCOE) in the autonomous case with 100% RE. Whilst the commercial sector with schools and hospitals also is important in studies about remote areas, larger industries and the transport sector are considered almost exclusively in case studies for municipalities, cities, islands or larger regions. In [72,73], industries are also considered in remote villages. However, they are referred to as *rural industries*, which probably corresponds more to the commercial sector of HIC in terms of demand structure. An interesting point is that in the cases where heat and industry were regarded, only a balanced autonomy is part of the analysis. This is probably due to the fact that, for example, high-temperature heat in industry can only be generated with specific RE plants and, in the completely autonomous case, would be associated with excessively high costs.

For residential, commercial and industrial sectors, the demand is usually known in advance in the studies. However, for the consideration of the transport sector several different approaches are applied. In [70,100,104,123] a fixed fuel demand for traditional vehicles is covered. In [51,52,116], electric vehicles (EVs) are considered within the EnergyPLAN model. In Dorotić et al. [51] all vehicles and ferries on the island of Korčula in Croatia are replaced by electrically powered alternatives. The EVs not only serve as batteries, but can also be used for vehicle-to-grid, i.e. feeding electricity from the EV battery into the grid. Šare et al. [116] analyse three scenarios for the municipality Dubrovnik in Croatia with different EV penetrations in 2020, 2030 and 2050. Krajačić et al. [79] and Oldenbroek et al. [99], on the other hand, included fuel cell vehicles in their ES analyses. In some scenarios in [79] the transport load is covered 100% by renewable hydrogen. None of the studies optimizes the number of electric or fuel cell vehicles.

### 4.3.    Feasibility and type of autonomy

In the reviewed literature, studies on completely autonomous ESs predominate with 110 (89%), whereas balanced LEA is only considered in 14 (11%) studies. The only study that analyses both cases seems to be Sameti and Haghighat [112], in which a net-zero energy district is investigated in three scenarios with grid connection and one as a stand-alone variant without grid connection.

Generally EA is feasible in the case studies. The only exception is the study by Alhamwi et al. [30] who do not obtain a feasible solution in their ES model and come to the conclusion that an off-grid city (165,000 inhabitants) is economically and technically not practicable. However, there are also other examples which do not come to a favourable result for LEA. Krajačić et al. [79] find that the cost of electricity for a *100% renewable island* is up to 15 times higher than the current (2009) electricity price. Furthermore, in Oldenbroek et al. [99] a 100% renewable supply can only be achieved if 20% of the vehicle fleet are fuel cell vehicles. Also Šare et al. [116] come to the conclusion that large storage capacities are necessary for a 100% renewable supply. Jenssen et al. [70] show that the available biomass potentials of a model municipality are sufficient for 100% power and heat supply, but not to replace transport fuel. All these examples have in common that they examine bigger regions, cities and islands as case studies in high-income countries in Europe.

In the studies on completely autonomous ESs, uncertainties due to disconnection from the grid infrastructures should play a very important role, since a non-optimal design of the ES cannot be compensated by imports. Therefore it is even more important to design these ESs robustly. There are several appropriate approaches in the studies. For example, in 39 (32%) studies a possible security of supply below 100% is implemented as a *loss of power supply probability* (LPSP). In most cases ([34–36,45,47,58,59,63,65,69,71–73,81,82,84,85,89–91,95–97,101,109,113,114,127,129]), the LPSP is modelled as a fixed value or results from other fixed values. Other studies ([37,61,68,83,98,106,115]) in which the LPSP is associated with weightings or penalty costs and thus integrated into the objective function of an optimization represent an improvement. In Hakimi et al. [61] different penalty costs were assumed for the residential, commercial and industrial sector. In future studies, the so-called *value of loss load* could be a suitable estimation of penalty costs. In Shivakumar et al. [388], for example, the value of loss load was calculated for households in all European Union member states. This data set with penalty costs based on the same methodology could make results of studies more comparable. As expected, the LPSP are rarely considered in HICs, as the inhabitants are accustomed to high security of supply. Four case studies in Canada [145], Sweden [37] and Hong Kong [85,134] are the only examples. However, for autonomous systems these LPSP become more relevant.

Further studies try to robustly design off-grid ESs by taking extreme conditions into account. In Petrakopoulou et al. [102,103], the plants of the ES are over-dimensioned and complementary technologies are used. In addition, the optimization model in Weinand et al. [124] considers extreme days on which demand is particularly high and no wind or solar radiation is present.

### 4.4. Time structure and pathway

In this section the time horizon (section 4.4.1), the chosen temporal resolution (section 4.4.2) and the pathway for the ES transition (section 4.4.3) are demonstrated.

#### 4.4.1. Time horizon

The time horizon in the case studies is usually chosen between 15 and 25 years, which represents an appropriate choice for estimating total discounted system costs or LCOEs for an ES. However, there are also variations upwards and downwards: Adamarola et el. [28] and Drydale et al. [52] even consider 35 and 45 years respectively. Jenssen et al. [70], Moeller et al. [94], Oldenbrock et al. [99], Østergaard and Lund [100] as well as Šare et al. [116] consider one year whereas Kandil et al. [74] use only a time horizon of 24 hours. For the latter study, a time horizon of 24 hours could be too short, even though only the operating costs of an autonomous ES are determined. At least an extreme day should have been considered for this analysis.

#### 4.4.2. Time resolution

The time resolution of models is of particular importance in studies on EA. This is especially true for completely autonomous ESs (cf. section 4.3). Non-optimal design of balanced energy autonomous systems could be compensated by imports from surrounding energy infrastructures. This does not apply for complete autonomy. Therefore, a particularly critical assessment is made when off-grid ESs are designed on the basis of an annual energy balance, as it seems in Stephen et al. [122]. Stephen et al. [122] investigate the residential and commercial energy supply for an off-grid Canadian aboriginal community. There are also other examples with a very rough time resolution, but these studies only consider balanced autonomy: Jenssen et al. [70] and Peura et al. [104] also conduct an annual balancing of energy (i.e. one time step) whereas the optimization model of Schmidt et al. [117] is based on two seasons (winter/summer) per year (i.e. two time steps).

In almost all studies (91, 74%) the time resolution is set to hours. There is only one study with a higher time resolution, namely Kötter et al. [78] with 15-minute time steps. Kötter et al. investigate the balanced EA of a region consisting of 17 sub-regions in Germany. However, it is not clear how many of the 15-minute time steps are used in the analysis. The robustness of results on completely autonomous ESs based on models with hourly resolution must at least be questioned. In these cases it is even more important to consider the methods explained in the previous section, such as LPSP or extreme conditions. In addition, ESs based on base-load capable technologies such as biomass can be considered more robust than those based only on volatile energy such as wind or photovoltaics (PV) (more on this in section 4.5). Usually all hours of a year are considered in the investigations with hourly resolution (59 of 91, 65%).

Overall it seems, however, that the number of time steps decreases with the complexity of a model, presumably in order to avoid computing time problems: the RE³ASON model based on public data uses only 288 [92] or 432 [124] time slices and the multi-objective optimizations of Fuentes-Cortés and Ponce-Ortega [54] or Yazdanpanah Jahromi et al. [125] use only 96 and 744 time slices respectively. Another example is the optimization with *multi-tier targets* (e. g. scenarios with different demands) according to the *World Bank Global Tracking Framework* by Fuso Nerini et al. [56], which comprises only 18 time steps per year. This is a general problem of ES analyses. However, as mentioned above, the number of time steps is more crucial in ES analyses including complete EA.

### 4.4.3. Pathway

EA projects are always associated with the objective that the ES will become energy autonomous in the medium to long-term future. This means that there will be a *transition* over several years. However, in almost all reviewed studies (115, 94%), *overnight* is chosen as the pathway, i.e. the new ES replaces the old one immidiateley and not during several years. This would correspond to an inaccurate calculation of total discounted system costs or LCOEs, as demands and costs may change during the considered time horizon. Only in Dorotić et al. [51], Drysdale et al. [52], Fuso Nerini et al. [56], Krajačić et al. [79], McKenna et al. [92] and Weinand et al. [124] was the pathway modeled as a transition. Dorotić et al. [51] seem to simulate at least every second year in EnergyPLAN from 2011 to 2030. The $CO_2$ emissions of the system are decreasing and the REs share is increasing until they reach their minimum (0% $CO_2$ emissions) or maximum (100% RE share) values in 2030. Drysdale et al. [52] also use the EnergyPLAN model. However, they seem to simulate only two years, 2016 and 2050. Fuso Nerini et al. [56] apply the system optimization model *OSeMOSYS* for a case study village in Timor Leste. The authors seem to optimize every year from 2010 until 2030. However, as mentioned above, for each year only 18 time steps are considered (six per day and three seasons per year), i.e. 360 time steps in total. Thereby the demand changes during the time horizon. For example, it is assumed, that the households reach the *target tier* in 2025. The target tier would be one of five tiers: for example the households would get access to general lighting, air circulation and television in *tier-2* or small appliances in *tier-3*. In the $H_2RES$ model in Krajačić et al. [79] every fifth year from 2005 until 2015 is simulated. The same applies to the RE³ASON model in McKenna et al. [92] and Weinand et al. [124] (time horizon from 2015 until 2030). In the two latter studies, however, the method is a mixed-integer linear optimization: all four years are optimized simultaneously, i.e. it is decided when which plant or measure will be installed. By considering the existing infrastructure (e.g. already installed PV modules), as in the RE³ASON model, models are enabled to consider a transition pathway.

### 4.5. Technologies

As already discussed in section 4.1, many case studies on energy autonomous remote rural areas deal with the reduction of diesel and the increase of REs in the system. As shown in Table 9, diesel, therefore, is the most frequently considered in the studies after PV, wind and stationary batteries. A total of 73 studies (59%) consider conventional generation technologies such as diesel generators and gas fired CHP / turbine plants in their ES analyses.

When classifying biomass CHP, hydropower plants, deep geothermal plants as well as conventional generation technologies as baseload-capable, 26 studies (21%) remain, in which only volatile generation technologies are considered. In 16 of these 26 studies, no long-term storage options such as hydrogen storage, pumped-hydro-storage or power-to-gas are considered. In such cases, it is essential to take account of uncertainties. In ten of the 16 studies [34–36,65,69,84,90,114,115,134], these uncertainties are at least addressed via LPSP and in another study [102] by including extreme conditions. Even more than in other studies, the results of the completely autonomous ESs in Al-Shetwi et al. [31], Khan et al. [75], Kim et al. [76] and Mas`ud [137] must therefore be questioned, in which only volatile energy technologies and no uncertainties are considered.

The fact that so few studies examine heating or cooling technologies (cf. Table 9) is related to the focus on electricity in the studies (cf. section 4.2.2). In addition, technologies that do not belong to the standard technologies such as PV or wind are primarily investigated in case studies in HIC in Europe. For example, the technologies deep geothermal energy, power-to-gas or district heating are analysed primarily in Germany (deep geothermal energy: [124]; power-to-gas: [78,94]; district heating: [70,124] ) or Denmark (deep geothermal energy (only heat): [100]; district heating: [52,100]), while unconventional vehicles such as EVs [51,52,116] or fuel cell vehicles [79,99,141] are examined primarily in case studies in Croatia. This suggests that the studies on remote rural areas are primarily concerned with the electrification of the area and not with the choice of optimal energy technologies. On the other hand, technologies such as deep geothermal energy (despite high potential in e.g. India or Sub-Saharan Africa [389]) are not relevant for these rather small regions (see section 4.2.1) due to high fixed costs [124].

*Table 9: Classification of the technologies included in the reviewed literature as well as the frequency of their consideration.*

| Category | Technology | No. of studies |
|---|---|---|
| Renewable electricity generation technologies | PV | 117 (95%) |
| | Wind (onshore) | 85 (69%) |
| | Biomass CHP | 39 (32%) |
| | Hydropower plant | 21 (17%) |
| | Concentrated solar power | 3 ( 2%) |
| | Deep geothermal plant | 2 ( 2%) |
| Heating / cooling technologies | District heating / cooling | 10 ( 8%) |
| | Heat pump | 7 ( 8%) |
| | Solar thermal collector | 6 ( 5%) |
| | Electric heater | 6 ( 5%) |
| Storage technologies | Stationary battery | 93 (76%) |
| | Hydrogen with fuel cell | 18 (11%) |
| | Thermal | 8 ( 7%) |
| | Pumped-hydro | 4 ( 3%) |
| | Power-to-Gas (methanisation) | 2 ( 2%) |
| Transport technologies | Electric vehicle (modelled with battery) | 3 ( 2%) |
| | Fuel cell vehicle | 3 ( 2%) |
| Conventional generation technologies | Diesel generator | 63 (51%) |
| | Gas fired CHP | 6 ( 5%) |
| | Gas turbine plant | 6 ( 5%) |

In summary, the studies on LEA investigate a wide range of technologies. However, for a robust design of an energy autonomous system based on REs, the combination of fluctuating and non-fluctuating generation technologies as well as different storage technologies could be advantageous. Some of these technologies that could be beneficial in a completely autonomous case, such as seasonal heat storage, have not yet been analysed. In general, the more diverse the technologies under consideration, the more economically or environmentally sustainable the ES could be designed. On the other hand, the complexity and computing time of ES models increases with the number of technologies. In any case, work still needs to be done in which a very broad range of technologies is considered and the optimum technologies for EA are identified. Based on the results of the reviewed studies, no definite trend towards the most economic technologies for achieving EA can yet be identified (cf. section 4.7).

### 4.6. Grid infrastructures

Grid infrastructures are rarely modeled in the studies. Heating grids are implemented only in [92,112,124], the electricity grid only in [30,92,94,124]. In [26,77] at least the costs for setting up the distribution network are taken into account. It is interesting to note that all four case studies that consider the electricity grid are located in Germany. In [92,124] the electricity and heating network is only represented in a simplified way by energy flows between districts. However, Weinand et al. [124] also contains a transferable approach for designing district heating networks in arbitrary municipalities. District heating systems are also designed in Sameti and Haghighat [112]. While in [124] the district heating network is modelled top-down for entire municipalities, [112] is better suited as a bottom-up application for districts for which exact building locations and energy demands are known. Therefore, depending on the application, the two studies offer possible approaches for future analyses.

In Moeller et al. [94], the capacities and connections of the electric transmission network between German regions are modeled. The analysis also examines whether the transmission capacities are sufficient, depending on the share of REs. In the FlexiGIS model in Alhamwi et al. [30], OpenStreetMap is used to obtain data on lines and substations of the distribution network in order to determine the optimal placement of a battery storage in an urban area. Unfortunately, power grid data is not yet completely included in OpenStreetMap and therefore this method is not usable for every case study.

Electricity grids are only considered in a simplified way in the papers. A promising approach for future studies could be the one of Morvaj et al. [156], in which the distribution network is modelled according to a linearized AC power flow approach. Of interest is the implementation of a binary modelling variable, which determines whether the distribution grid needs to be upgraded, depending on the amount and type of REs added to the ES. In the case of an upgrade, the expansion of REs would involve additional costs. However, the study uses available grid data of the *IEEE European Low Voltage Test Feeder case* [390]. Since this grid data is not available for arbitrary case studies, the grid capacities would have to be estimated.

### 4.7. Costs

Section 4.1 has already shown that the ESs in the literature were mostly designed on the basis of cost minimization. Therefore, a comparison of these costs is reasonable. In 83 (68%) of the 123 studies, the LCOEs for autonomous ESs were stated (cf. Figure 3). For Figure 3, the LCOEs from the studies were adjusted according to inflation until 2019 [391] and converted into $/kWh using the average exchange rates [392–394] in the year of the respective publication. As all but one of the 83 studies consider the residential sector, the household electricity price (import from grid) in the different countries is shown for comparison [395]. For eight countries (e.g. Ethiopia [66] or Yemen [31]), the household electricity price could not be found.

The mean LCOEs amount to 0.41 $/kWh (black dotted line in Figure 3 and Table 10). Consequently, the costs in the studies of Khan et al. [75] and Hosseini et al. [65] are nearly average. The LCOEs in an ES with 100% RE (case studies with 100% RE as well as with 100% RE and a security of supply < 100% in Figure 3 and Table 10) are on average 0.42 $/kWh (0.37 $/kWh without the outlier in Askari and Ameri [34]) whereas in an ES with conventional energy 0.39 $/kWh (0.36 $/kWh without the outlier in Shezan et al. [121]) is achieved. As expected, in studies considering both cases, the LCOEs of 100% RE systems are higher than of conventional ESs. Likewise, the LCOEs decrease if cases with security of supply below 100% are considered. Furthermore, the household electricity prices are lower than the LCOEs for the autonomous system for almost every study. The only exceptions include case studies in countries with above-average electricity prices (Australia [40], Croatia [51] and Germany [78,94]), or the two studies with the lowest LCOEs [106,109] in Figure 3.

*Table 10: Mean levelized cost of electricity in the 83 studies depending on the characteristics of the energy systems. The LCOEs were adjusted according to inflation until 2019 [391] and converted into $/kWh using the average exchange rates [392-394] in the year of the respective publication.*

| Characteristics of energy system | Number of studies | Mean LCOE [$/kWh] | Highest outlier [$/kWh] and respective study | Mean LCOE without highest outlier [$/kWh] |
|---|---|---|---|---|
| All energy systems | 83 | 0.41 | 3.27 (Askari and Ameri [34]) | 0.38 |
| Including conventional energy | 48 | 0.37 | 0.95 (Bekele and Palm [66]) | 0.36 |
| Including conventional energy (security of supply < 100%) | 5 | 0.61 | 2.01 (Shezan et al. [121]) | 0.26 |
| 100% RE | 37 | 0.41 | 1.57 (Li [82]) | 0.38 |
| 100% RE (security of supply < 100%) | 18 | 0.43 | 3.27 (Askari and Ameri [34]) | 0.27 |

18 out of the first 20 upward outliers in Figure 3 apply the HOMER model (for the other two, the applied model cannot be found in the article), i.e. a non-optimal design of the ES could be responsible for the high LCOEs (cf. section 3.1). In the following, the three studies from Figure 3 with the highest upward outliers in the LCOEs are discussed (up to the study Li [82] with 1.57 $/kWh on average). The highest LCOEs in the study by Askari and Ameri in 2009 [34] are caused by the high inflation in Iran between 2009 and 2019 (+364%). As a result, the costs are adjusted from originally 0.75 $/kWh to 3.27 $/kWh. The study by Shezan et al. [121] with the high LCOEs of 2.01 $/kWh needs further investigation, especially since an unmet load of 0.01% is considered here, which should reduce the LCOEs. Unfortunately, by analysing the study it is not really possible to determine how the HOMER model achieves the high LCOEs for the most economic ES with PV, wind, diesel generator and stationary battery. However, surprisingly, a figure in the study shows more realistic HOMER results with LCOEs of 0.62 $/kWh, but these LCOEs are not discussed further in the text. In Li [82] the predefined design of the ES seems to lead to the high LCOEs of 1.54 $/kWh. In fact, an ES with 500 kW of PV and 9.1 MWh of stationary batteries is assumed for 100 households in China. This ES appears to be oversized, which again demonstrates the need for a good understanding of the ES when designing ESs with HOMER (cf. section 4.1). In addition, the PV-battery system is compared with a PV-battery-fuel cell system, but the capacities of PV and battery are not changed. Thus, it is obvious that the PV-battery system leads to lower costs.

The reason for the low LCOEs for energy autonomous systems of the downward outliers is more difficult to determine and would require an in-depth analysis in a separate study. In the case study in Rajanna and Saini [106], for example, there is great potential for baseload hydropower and bioenergy, which could be related to the low LCOEs of 0.07 $/kWh. Examining further studies for reasons related to LCOE would be beyond the scope of this literature review. However, the detailed table in the supplementary Excel file could be used in further studies to investigate the dependencies of the LCOEs on the characteristics of the studies, e.g. through cluster or regression analyses. The supplementary Excel File and Figure 3 are useful as a basis for evaluating future studies about LEA. If, for example, the LCOEs deviate as much from the average of 0.41 $/kWh in future studies as in Li [82], the applied methods and results have to be further investigated.

Since most of the studies calculate LCOEs, these figures are very suitable for comparing the results of LEA studies. However, another cost parameter that is particularly relevant for the inhabitants who have to pay for the costs of the autonomous ES is rarely shown in the studies: total costs per inhabitant. These are only shown in Jenssen et al. [70] (1.4 – 2.3 k$), Schmidt et al. [117] (220 $/a more than in the reference scenario without autonomy) and Weinand et al. [124] (21.0 – 54.8 k$). However, the cost per inhabitant should be included in all future studies in order to assess the feasibility of the EA project.

In some studies, the costs of the completely autonomous ES are compared with the costs of grid connection [33,87,97,101,111,131]. In all examples, the grid connection scenario turns out to be less economical. This is due to the fact that these studies only consider remote areas that are far away from the nearest grid connection point. The break-even points for the distance from which the network connection would be worthwhile are calculated in [87,101,111,131]. This is also related to the question of the optimal degree of centralization as well as the optimal size of energy autonomous municipalities (cf. section 5).

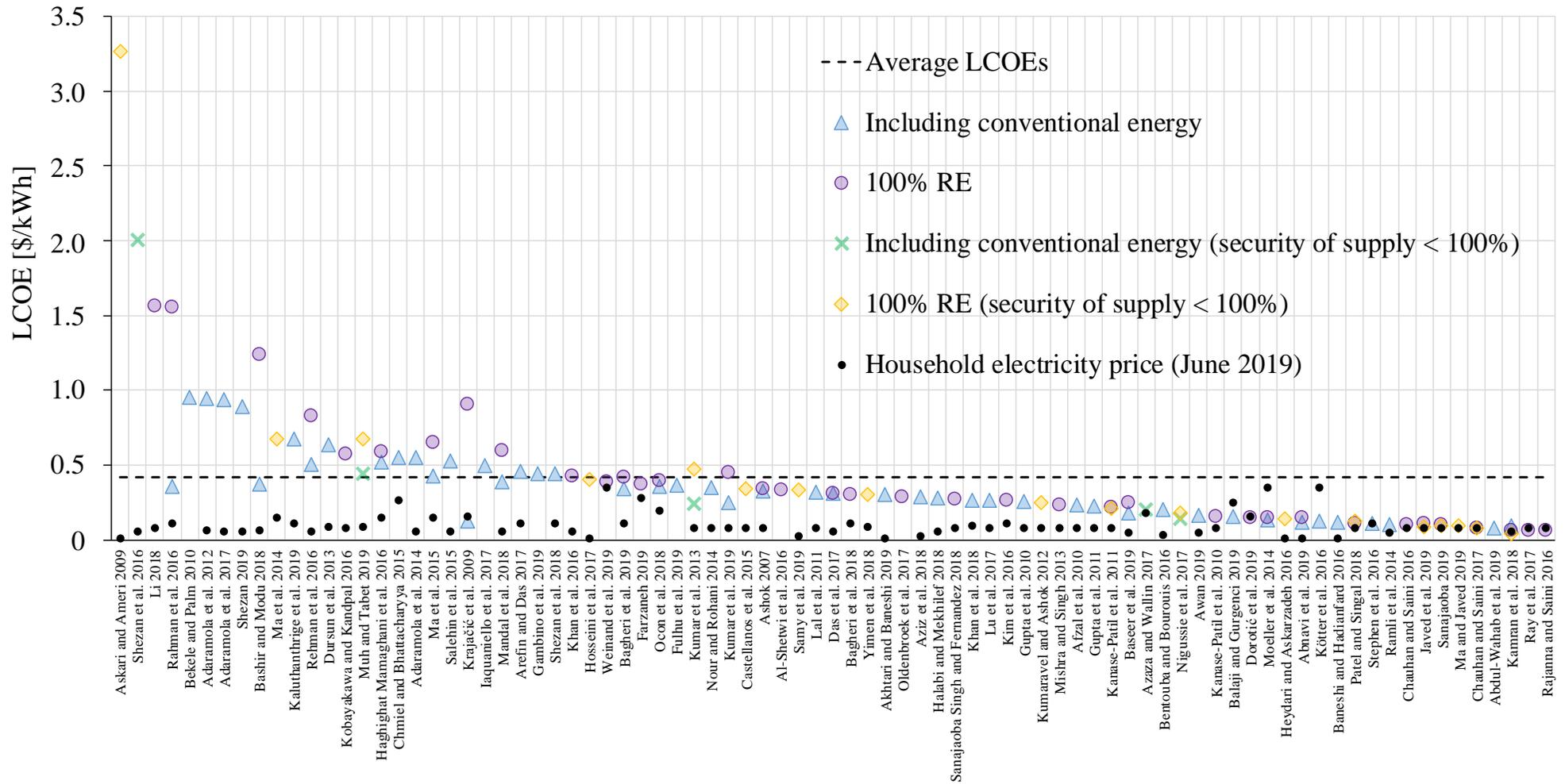

*Figure 3: LCOEs of the energy autonomous case studies in the literature. The studies are sorted by mean LCOEs, from high to low. The LCOEs were adjusted according to inflation until 2019 [391] and converted into $/kWh using the average exchange rates [392-394] in the year of the respective publication. The household electricity price (black dots) in the different countries is shown for comparison [395]. The black dotted line shows the average value (0.41 $/kWh).*

## 5. Critical appraisal of energy autonomy studies

In this section, some of the key findings of section 4 are subjected to further critical evaluation. Firstly, though it is difficult to generalize across all 123 reviewed studies, some emerging trends may be highlighted:

- Mostly conventional/established technologies are analysed, with less attention paid to emerging but potentially game-changing technologies such as deep geothermal and fuel cell vehicles;
- The sectoral focus is on residential, with much less consideration of industrial and transportation sectors;
- Network infrastructure is rarely considered, including electricity, gas and heat/cooling;
- Only a minority of studies account for the existing infrastructure as well as the transition from this state to some improved future state along a transition pathway;
- Most studies focus on complete EA, with some (12%) dealing with balanced EA.

The time resolution of most studies is hourly, with one study going into more detail at the 15-minute level. For long-term planning purposes, the hourly resolution is suitable, but it should also be combined with more detailed analyses and/or information reduction techniques to generate time typologies and synthetic time slices. There is an established stream of research focussing on the most suitable/required time resolution for specific research questions, for ESs with significant renewable generation fractions (e.g. [396,397]). Studies addressing regional and LEA could also benefit from exploring such approaches. Otherwise there is a danger that systems are incorrectly dimensioned and are inadequate to ensure supply security in times of peak demand, which for autonomous/off-grid systems is potentially critical.

There is also a lack of attention paid to non-economic and non-technical criteria in studies of EA at the local scale. Indeed, economic criteria are arguably the most important, but although they are necessary, they are not sufficient. In section 1, among others, it has already been highlighted that certain areas are very far from the national grid and therefore a stand-alone system is appropriate. In this case, the focus on costs as a target criterion is justified, however, the comparison of complete autonomy with the grid connection case should always be demonstrated. Otherwise, in future analyses of EA more criteria than costs need to be considered. There are other important reasons for municipalities to become energy autonomous besides costs, such as increased environmental awareness (cf. section 1). Furthermore, a comparison with the electricity price in section 4.7 shows that EA usually leads to significantly higher costs, and thus from a cost perspective there is no potential for EA.

In terms of the modelling approaches employed for highly renewable, autonomous ESs there is a clear dominance of linear programming (i.e. optimization) and simulation (i.e. dispatch rules for energy balancing), from a central planner perspective. It is encouraging to find that many researchers are also capturing the non-economic criteria such as health, noise, water and acceptance issues (cf. Table 7). However, these contributions are still in the minority of those reviewed here, and the overwhelming majority do not satisfactorily reflect the true complexity encountered in real-world energy transitions. It is common practice to leave stakeholder roles outside the scope of the studies or models and to calculate optimal autonomy transition pathways with a centralized planning approach under the selected objective of technical feasibility and economic viability. However, local ESs are complex socio-technical systems consisting of different decision-making entities and technological artefacts governed by energy policy in a multi-level institutional space [11]. Social relationships among the stakeholders represent a major driver or barrier as also stated by previous reviews [4]. In this context, adoption behaviour approaches are useful to understand the types of barriers that exist for new technologies, and what kind of policies are important to increase diffusion. As a consequence, realising the potential of LEA is not simply a question of technical realities but also a question of individual behaviour and group dynamics. Relevant local stakeholders as households or communities, energy producers, energy suppliers, service providers, as well as local policy-makers are inter-dependent in the realization of their goals. Future system models need to include the heterogeneous roles different stakeholders play in an existing local environment and the resulting impact their decision making might have. One possible solution could be the extension of the presented techno-economic modelling approaches with the help of socio-economic modelling approaches as agent-based models [398,399] or system dynamics models [400].

Many different spatial scales (e.g. number of households) have been considered in the case studies (cf. section 4.2.1). Whilst the optimal spatial size of an energy autonomous region has not yet been identified, it is interesting to note that the mean number of households and inhabitants in energy-autonomous regions is 340 and 18,200, respectively (based on the 56 and 49 articles containing this information, respectively). This is strongly linked to the question of the optimal degree of centralisation [401]. Concerning demand and consumption sectors there is also some potential for improvement (cf. section 4.2). The demand product water, for example, which is strongly linked to energy, or the consumption sector industry is only very rarely taken into account. For the latter, transferable methods to determine the energy demand and load profile of industries in arbitrary regions could facilitate its implementation. The transport sector is also almost completely ignored in studies about LEA. A particularly interesting approach would be to optimize the number and use of electric or fuel cell vehicles. In general, however, all sectors should be taken into account, especially when estimating the impact of one or many energy autonomous regions on the surrounding ES.

Validation is challenging in the context of local ESs, which might explain why very few existing studies attempt to do this. Often detailed data on the existing ES is lacking and validation for some hypothetical future scenario is obviously not meaningful. Model design and data assumptions of studies used to gain insights to form the decision making should be transparent and accessible. This not only allows independent review of various stakeholders but also the complete reproducibility of the results [402]. Whilst there is a strong trend towards open-source models and data within the wider ES modelling field [403], of the reviewed models, none appear to be fully open source, with Homer and EnergyPlan only being open access. Hence LEA studies could increase efforts to publicly release data and system models as well as assumptions and results interpretation, in order that diverse affected stakeholders are able to participate in the decision-making process [402,404].

In the future, the increasing availability of large(r) amounts of (more) open data should facilitate the implementation of data-driven methodologies. These approaches are already providing helpful insights into different aspects of energy production and demand. For example, machine learning approaches are capable of capturing non-linear and complicated relationships, and could be implemented more often [405]. In future EA case studies, such approaches could be used to detect existing renewable energy plants [406], predict renewable energy generation [407] or predict energy demands [408,409]. Data availability can therefore not only facilitate the validation of results, but also create novel insights through the application of innovative methods. Improved data quality should help to increase the accuracy of energy system models for decentralized energy systems.

In addition, the focus on the LCOEs as a benchmark for highly-RE systems could provide potentially misleading results. Whilst the LCOE is a good first indicator of the generation costs and allows comparisons across technology, it is noted for neglecting the additional costs of integrating non-dispatchable renewable technologies into the ES. Three additional cost components should be considered, if the true *system LCOEs* of RE technologies are to be considered [410,411]:

- **Profiling costs**, related to the requirement for the dispatchable generation technologies to meet the residual load;
- **Balancing costs**, related to the deviation between forecast and actual non-dispatchable renewable generation; and
- **Grid costs**, related to addition grid reinforcement and extension (at all voltage levels) required to connect renewable generators to the network.

Attempts have been made to consider these cost components in the context of large-scale (e.g. national) ES analyses (e.g. [412–414]). However, at the regional and municipal scale, as demonstrated by this review, they are typically not included. This is despite the fact that, when considering balanced autonomy, these effects on the surrounding ES are of particular importance. Balanced EA and the associated increasing feed-in by renewables could make network expansion even more essential and also make new allocation systems for grid fees necessary [401]. The result could be economic inefficiencies compared to the established system of centralised generation, transmission and distribution [415]. However, possible ES impacts were not considered in any of the 14 studies on balanced autonomy. Hence there is a need for further research to address these and the above mentioned deficits.

## 6. Summary and conclusions

Research attention on decentralized autonomous energy systems has increased exponentially in the past three decades, as demonstrated by the absolute number of publications and the share of these studies in the corpus of energy system modelling literature. This paper shows the status quo and future modelling needs for research on local autonomous energy systems. A total of 359 studies are roughly investigated, of which a subset of 123 in detail. The studies are assessed with respect to the characteristics of their methodology and applications, in order to derive common trends and insights.

The results show that most case studies were conducted in the middle-income countries India, Iran and China as well as the high-income country Germany. In the middle-income country studies, mostly remote rural areas without electricity network connection are considered, whereas in high-income countries the case studies are much more diverse and also include cities and islands. In addition, most studies only focus on the residential sector and the supply of electricity. A wide range of technologies has already been covered in the literature, including less common technologies such as power-to-gas and fuel cell vehicles. However, the network infrastructure is rarely considered. The levelized costs of electricity for local autonomous energy systems in 83 case studies amount to 0.41 $/kWh on average. Thereby, studies are identified in which the resulting costs should be questioned, as they deviate strongly from the average.

In terms of the employed methodology, most of the reviewed literature reports an optimization or simulation approach, with a central planner perspective. They typically employ a time resolution of one hour, but for some studies also increase this to 15-minute resolution. Whilst it is commendable that some of the studies also consider non-economic criteria such as social and environmental aspects, neither the system-level impacts nor the diverse stakeholders are included in most works. Furthermore, there is a general lack of transparency across most reviewed literature, meaning that neither open data nor open models are widely applied to local energy systems.

Hence, future research should focus on the following methodological innovations. Other perspectives than that of a central planner and other target criteria than costs should be included. This could contribute to the realizability of the case study results. System impacts of many local autonomous energy systems have not yet been investigated, which could make new distribution systems and grid fees necessary. Complete autonomous energy systems in particular must be robustly designed, for example by analysing the value of lost load and whether a security of supply below 100% is acceptable for consumers of the case study, preferably using penalty costs for unmet load in the target function. In addition, extreme conditions such as extreme days with low solar radiation or wind should be considered and the temporal resolution should be higher than the usually used hourly resolution. Finally, methodologies should be developed which can involve local stakeholders in the modelling process and thus consider their preferences relating to their future energy system.

**Acknowledgements**

Fabian Scheller receives funding from the European Union's Horizon 2020 research and innovation programme under the Marie Sklodowska-Curie grant agreement no. 713683 (COFUNDfellowsDTU). Russell McKenna and Fabian Scheller kindly acknowledge the financial support of the FlexSUS Project (Project nbr. 91352), which has received funding in the framework of the joint programming initiative ERA-Net Smart Energy Systems' focus initiative Integrated, Regional Energy Systems, with support from the European Union's Horizon 2020 research and innovation programme under grant agreement No 775970. Russell McKenna also gratefully acknowledges the support of the Smart City Accelerator project. The usual disclaimer applies.